\documentclass[a4paper,11pt]{article}
\usepackage{latexsym}
\usepackage{amsmath}
\usepackage{amssymb}
\usepackage{graphicx}
\usepackage{wrapfig}
\begin{document}

\noindent
SAGA-HE-217-04

\baselineskip=7mm

%\begin{flushleft}

\bigskip

\centerline{\bf Auxiliary field method at finite temperature and/or finite density}

\centerline{Hiroaki Kouno}
\centerline{\it Department of Physics, Saga University, Saga 840-8502, Japan}

\baselineskip=3mm

~

\small
\centerline{\bf Abstract}
We discuss the relation between the effective vector meson mass and equation of state (EOS) for nuclear matter. 
In the mean field approximation, the EOS becomes softer due to the reduction of the effective $\omega$-meson mass, if we assume that the $\omega$-meson mean field is proportional to the baryon density. 
We examine the assumption by using the auxiliary field method at finite temperature and /or density. 

~

\noindent
{\bf 1 Introduction}

Medium effects on the vector meson masses are much interested in the hadron and nuclear physics. [1] 
Because of its short life time, the reduction of the $\rho$-meson mass is expected to be a signal of the hot and dense matter which may be produced in the high-energy heavy ion collisions. [2] 
On the other hand, the $\omega$-meson is important for the nuclear structure. 
It is reported that the reduction of the effective vector meson mass makes the nuclear matter EOS stiffer. [3] 

In this paper, we discuss the relation among the effective vector meson masses, the effective vector meson-nucleon coupling and the nuclear EOS, using the generalized mean field theory. [4,5] 
We show that, if we assume that the $\omega$ ($\rho$)-meson mean field is proportional to the baryon (isovector) density, the effective $\omega$ ($\rho$)-nucleon coupling also becomes smaller as the effective $\omega$ ($\rho$)-meson masses becomes smaller and the EOS becomes softer. [5] 
We examine the assumption by using the auxiliary field method [6] at finite temperature and/or finite density. 
It is shown that, in the simple model with four fermion interactions, the value of the $\omega$-meson mean field is exactly proportional to the baryon density. 

~

\noindent
{\bf 2 Effective meson mass and effective meson-nucleon coupling} 

In the generalized mean field theory, [4,5] 
the effective couplings $\hat{g}_{\sigma,\omega,\rho}$ and the effective meson masses $m_{\sigma,\omega,\rho}^*$ are defined by 
%%%%%%%%
\begin{eqnarray}
\hat{g}_\sigma&=&-{\partial \Sigma_{\rm v}\over{\partial \sigma}}+{m^*\over{E_{\rm F}^*}}{\partial \Sigma_{\rm s}\over{\partial \sigma}},~
\hat{g}_\omega =-{\partial \Sigma_{\rm v}\over{\partial \omega}}+{m^*\over{E_{\rm F}^*}}{\partial \Sigma_{\rm s}\over{\partial \omega}},~
\hat{g}_\rho =-{\partial \Sigma_{\rm v}\over{\partial \rho}}+{m^*\over{E_{\rm F}^*}}{\partial \Sigma_{\rm s}\over{\partial \rho}},
\nonumber\\
{m_{\sigma}^*}^2&=&{\partial^2 \epsilon\over{\partial \sigma^2}},
~{m_{\omega}^*}^2=-{\partial^2 \epsilon\over{\partial \omega^2}}
~{\rm and}
~{m_{\rho}^*}^2=-{\partial^2 \epsilon\over{\partial \rho^2}}
;~~ E_{\rm F}^*=\sqrt{k_{\rm F}^2+{m^*}^2}, 
\label{eq:E1}
\end{eqnarray}
%%%%%%%%
where $\sigma$, $\omega$, $\rho$, $\Sigma_{\rm s}$, $\Sigma_{\rm v}$, $k_{\rm F}$, $m^*$ and $\epsilon$ are the $\sigma$-meson mean field, the $\omega$-meson mean field, the $\rho$-meson mean field, the scalar self-energy for the nucleon, the vector self-energy for the nucleon, the Fermi momentum, the effective nucleon mass and the energy density for the nuclear matter, respectively. 
If there is no mixing element in the effective meson-mass matrix, [4,5] 
the first derivative of the pressure $P$ for the symmetric nuclear matter with respect to the baryon density $\rho_{\rm B}$ is given by
%%%%%%%%
\begin{eqnarray}
{dP\over{d\rho_{\rm B}}}&=&\left({k_{\rm F}^2\over{3\rho_{\rm B}E_{\rm F}^*}}+{\hat{g}_\omega^2\over{{m_\omega^*}^2}}+{\hat{g}_\rho^2\over{{m_\rho^*}^2}}-{\hat{g}_\sigma^2\over{{m_\sigma^*}^2}}{{m^*}^2\over{{E_{\rm F}^*}^2}}\right)\rho_{\rm B}. 
\label{eq:E2}
\end{eqnarray}
%%%%%%%%
If we assume that the value of the $\omega$-meson mean field is proportional to the baryon density, $\hat{g}_\omega$ is related to $m_\omega^*$ by the relation %%%%%%%%%
\begin{eqnarray}
{\hat{g}_\omega\over{g_\omega}}={{m_\omega^*}^2\over{m_\omega^2}}, 
\label{eq:E3} 
\end{eqnarray}
%%%%%%%%%
where $g_\omega$ and $m_\omega$ are the $\omega$-nucleon coupling and the $\omega$-meson mass at zero density, respectively. [5] 

Similarly, if the $\rho$-meson mean field is proportional to the isovector density $\rho_3$, the effective coupling $\tilde{g}_\rho$ for the asymmetric nuclear matter is related with $m_\rho^*$ by the relation 
%%%%%%%%%
\begin{eqnarray}
{\tilde{g}_\rho\over{g_\rho}}={{m_\rho^*}^2\over{m_\rho^2}}, 
\label{eq:E4} 
\end{eqnarray}
%%%%%%%%%
where $g_\rho$ and $m_\rho$ are the $\rho$-nucleon coupling and the $\rho$-meson mass at zero density, respectively. [5] 

If the values of the vector meson fields are proportional to the corresponding baryonic currents, the mixing elements of the effective meson mass matrix vanish and Eq. (\ref{eq:E2}) holds true. 
Putting Eq. (\ref{eq:E3}) into Eq. (\ref{eq:E2}), we obtain 
%%%%%%%%
\begin{eqnarray}
{dP\over{d\rho_{\rm B}}}&=&\left({k_{\rm F}^2\over{3\rho_{\rm B}E_{\rm F}^*}}+{\hat{g}_\omega g_\omega\over{{m_\omega}^2}}+{\hat{g}_\rho^2\over{{m_\rho^*}^2}}-{\hat{g}_\sigma^2\over{{m_\sigma^*}^2}}{{m^*}^2\over{{E_{\rm F}^*}^2}}\right)\rho_{\rm B}. 
\label{eq:E5} 
\end{eqnarray}
%%%%%%%%
If $m_\omega^*$ decreases, $\hat{g}_\omega$ also decreases according to Eq. (\ref{eq:E3}) and ${dP\over{d\rho_{\rm B}}}$ becomes smaller according to Eq. (\ref{eq:E5}). 
Therefore, the EOS becomes softer. 
On the other hand, due to the equation (\ref{eq:E4}), the EOS for the asymmetric nuclear matter becomes softer if the effective $\rho$-meson mass becomes smaller. [5] 

~

\noindent
{\bf 3 Auxiliary field method}

In the previous section, we assume that the value of the $\omega$-meson mean field is proportional to the baryon density. 
In view point of quark ($q$) physics, the assumption might be natural. [7] 
In this section, we examine the assumption, using the four fermion interaction model and the auxiliary field method [6] at finite temperature ($T$) and/or finite density. 

We start with the following generating function with finite source $J_\mu$ for the vector current $\bar{q}\gamma_\mu q$. 
%%%%%%%%
\begin{eqnarray}
Z(J)&=&\int d\bar{q}dq \exp{\left( -
\int_{\beta V} d^4x \left\{\bar{q}(\partial_\mu\gamma_\mu
-J_\mu\gamma_\mu)q -{\lambda\over{2}}(\bar{q}\gamma_\mu q )^2\right\}\right)}, 
\nonumber\\
\label{eq:E6}
\end{eqnarray}
%%%%%%%%
where $\beta=1/T$ and $V$ is the three dimensional volume. 
Inserting the identity for the auxiliary field $\Omega_\mu$
%%%%%%%%
\begin{eqnarray}
1&=&\int d\Omega\exp{\left\{-{1\over{2\lambda}}\int_{\beta V} d^4x\left(-\Omega_\mu +\lambda\bar{q}\gamma_\mu q\right)^2\right\}}
\label{eq:E7}
\end{eqnarray}
%%%%%%%%
into Eq. (\ref{eq:E6}), we obtain 
%%%%%%%%
\begin{eqnarray}
Z(J)&=&\int{d\bar{q}}{dq}{d\Omega} \exp{\left(\int_{\beta V} d^4x\left\{-{1\over{2\lambda}}\Omega_\mu^2
-\bar{q}(\partial_\mu\gamma_\mu
-(\Omega_\mu+J_\mu)\gamma_\mu)q \right\}\right)}
\label{eq:E8}
\end{eqnarray}
%%%%%%%%
If we define ${\tilde{\Omega}}_\mu =\Omega_\mu +J_\mu$, 
we obtain 
%%%%%%%%
\begin{eqnarray}
Z(J)=\int{d\bar{q}}{dq}{d\tilde{\Omega}} \exp{\left(\int_{\beta V} d^4x\left\{-{1\over{2\lambda}}\tilde{\Omega}_\mu^2+{1\over{\lambda}}\tilde{\Omega}_\mu J_\mu\right. \right. }
\nonumber\\
{\left.\left. -\bar{q}(\partial_\mu\gamma_\mu
-\tilde{\Omega}_\mu\gamma_\mu)q-{1\over{2\lambda}}J_\mu^2\right\}\right)}.
\label{eq:E9}
\end{eqnarray}
%%%%%%%%
Differentiating the logarithms of Eqs. (\ref{eq:E6}) and (\ref{eq:E9}) with respect to $J_\mu$, 
we obtain
%%%%%%%
\begin{eqnarray}
{3g_\omega\over{\lambda}}\omega_\mu&\equiv& {1\over{\lambda}}<\Omega_\mu>
={1\over{\lambda}}<\tilde{\Omega}_\mu>-{1\over{\lambda}}J_\mu
=<\bar{q}\gamma_\mu q >
\label{eq:E10}
\end{eqnarray}
%%%%%%%
Putting $J_i=0$ ($i=1,2,3$), $\omega =\omega_0$ and $\lambda =g_\omega^2/m_\omega^2$, we obtain, 
%%%%%%%
\begin{eqnarray}
\omega ={g_\omega\over{3m_\omega^2}}<\bar{q}\gamma_0 q>={g_\omega\over{m_\omega^2}}\rho_{\rm B}
\label{eq:E11}
\end{eqnarray}
%%%%%%%
Therefore, the $\omega$-meson mean field is proportional to the baryon density $\rho_{\rm B}$. 

~

\noindent
{\bf 4 Summary and discussions}

In summary, we have discussed the relation between the effective vector meson masses and equation of state (EOS) for nuclear matter in the framework of the generalized mean field theory. 
We have shown that, if we assume that the $\omega$ ($\rho$)-meson mean field is proportional to the baryon (isovector) density, the effective $\omega$ ($\rho$) -nucleon coupling also becomes smaller as the effective $\omega$ ($\rho$)-meson masses becomes smaller and the EOS becomes softer. 

We examine the assumption by using the auxiliary field method at finite temperature and/or finite density. 
In the simple model with four fermion interactions, the mean field of the $\omega$-meson, which is composed of quark and anti-quark, is exactly proportional to the baryon density. 
However, if we use the simplest mean field approximation, the meson-nucleon coupling does not have the density dependence for this simplest model. 
Therefore, it is needed to generalize our analysis beyond the mean field approximation or to generalize our auxiliary field method to more complex models with many fermion interaction. 
These works are now in progress. 

~

\centerline{\bf Acknowledgement}

The author would like to thank Prof. T. Kunihiro for useful discussions and suggestions. 
The author would also like to thank T. Sakaguchi, K. Tuchitani and Y. Horinouchi for the collaboration on the subject discussed in this work. 

\begin{flushleft}

{\bf References}

%%%%%%%%%%%%%%%%%%%%%%%%%%%%%%%%%%%%%%%%%%%%%%%%%%%%%%%%%%%%%%%%%%%%%%%%%%%
[1] G.E. Brown and M. Rho, Phys. Rev. Lett., {\bf 27} (1991) 2720:  
%%%%%%%%%%%%%%%%%%%%%%%%%%%%%%%%%%%%%%%%%%%%%%%%%%%%%%%%%%%%%%%%%%%%%%%%%%%
%%%%%%%%%%%%%%%%%%%%%%%%%%%%%%%%%%%%%%%%%%%%%%%%%%%%%%%%%%%%%%%%%%%%%%%%%%%
T. Hatsuda and S. H. Lee, Phys. Rev. {\bf C46} (1992) 46; 
%%%%%%%%%%%%%%%%%%%%%%%%%%%%%%%%%%%%%%%%%%%%%%%%%%%%%%%%%%%%%%%%%%%%%%%%%%%
%%%%%%%%%%%%%%%%%%%%%%%%%%%%%%%%%%%%%%%%%%%%%%%%%%%%%%%%%%%%%%%%%%%%%%%%%%%
T. Hatsuda and T. Kunihiro, Phys. Rep. {\bf 247} (1994) 221.  
%%%%%%%%%%%%%%%%%%%%%%%%%%%%%%%%%%%%%%%%%%%%%%%%%%%%%%%%%%%%%%%%%%%%%%%%%%%

%%%%%%%%%%%%%%%%%%%%%%%%%%%%%%%%%%%%%%%%%%%%%%%%%%%%%%%%%%%%%%%%%%%%%%%%%%%
[2] See, e.g., I. Tserruya, preprint nucl-ex/0204012.  
%%%%%%%%%%%%%%%%%%%%%%%%%%%%%%%%%%%%%%%%%%%%%%%%%%%%%%%%%%%%%%%%%%%%%%%%%%%

%%%%%%%%%%%%%%%%%%%%%%%%%%%%%%%%%%%%%%%%%%%%%%%%%%%%%%%%%%%%%%%%%%%%%%%%%%%%
[3] F. Weber, Gy. Wolf, T. Maruyama and S. Chiba, preprint nucl-th/0202071; 
%%%%%%%%%%%%%%%%%%%%%%%%%%%%%%%%%%%%%%%%%%%%%%%%%%%%%%%%%%%%%%%%%%%%%%%%%%%
%%%%%%%%%%%%%%%%%%%%%%%%%%%%%%%%%%%%%%%%%%%%%%%%%%%%%%%%%%%%%%%%%%%%%%%%%%%
C.H. Hyun, M.H. Kim and S.W. Hong, preprint nucl-th/0308053. 
%%%%%%%%%%%%%%%%%%%%%%%%%%%%%%%%%%%%%%%%%%%%%%%%%%%%%%%%%%%%%%%%%%%%%%%%%%%

%%%%%%%%%%%%%%%%%%%%%%%%%%%%%%%%%%%%%%%%%%%%%%%%%%%%%%%%%%%%%%%%%%%%%%%%%%%
[4] K. Tuchitani et al., Int. J. Mod. Phys. {\bf E10} (2001) 245. 
%%%%%%%%%%%%%%%%%%%%%%%%%%%%%%%%%%%%%%%%%%%%%%%%%%%%%%%%%%%%%%%%%%%%%%%%%%%

%%%%%%%%%%%%%%%%%%%%%%%%%%%%%%%%%%%%%%%%%%%%%%%%%%%%%%%%%%%%%%%%%%%%%%%%%%%
[5] H. Kouno et al., preprint, nucl-th/0405022; K. Tuchitani et al., preprint, nucl-th/0407004. 
%%%%%%%%%%%%%%%%%%%%%%%%%%%%%%%%%%%%%%%%%%%%%%%%%%%%%%%%%%%%%%%%%%%%%%%%%%%

%%%%%%%%%%%%%%%%%%%%%%%%%%%%%%%%%%%%%%%%%%%%%%%%%%%%%%%%%%%%%%%%%%%%%%%%%%%
[6] 
T. Kashiwa and T. Sakaguchi, Phys. Rev. {\bf D68} (2003) 589 , and references therein. 
%%%%%%%%%%%%%%%%%%%%%%%%%%%%%%%%%%%%%%%%%%%%%%%%%%%%%%%%%%%%%%%%%%%%%%%%%%%

%%%%%%%%%%%%%%%%%%%%%%%%%%%%%%%%%%%%%%%%%%%%%%%%%%%%%%%%%%%%%%%%%%%%%%%%%%%
[7] T. Kunihiro, private communication. 
%%%%%%%%%%%%%%%%%%%%%%%%%%%%%%%%%%%%%%%%%%%%%%%%%%%%%%%%%%%%%%%%%%%%%%%%%%%

\end{flushleft} 

\end{document}